\documentclass[conference]{IEEEtran}
\IEEEoverridecommandlockouts
\usepackage{cite}
\usepackage{amsmath,amssymb,amsfonts}
\usepackage{algorithmic}
\usepackage{graphicx}
\usepackage{textcomp}
\usepackage{xcolor}
\usepackage{flushend}
\usepackage[hidelinks]{hyperref}

\usepackage{float}
\def\BibTeX{{\rm B\kern-.05em{\sc i\kern-.025em b}\kern-.08em
    T\kern-.1667em\lower.7ex\hbox{E}\kern-.125emX}}
\begin{document}

\title{Workload Characterization for Branch Predictability\\
\thanks{{*}Note: This manuscript is an archival version of work conducted as part of the author's 2020 Master's thesis at Department of ECEN, Texas A\&M University, College Station, TX under the supervision of Professors Paul Gratz and Daniel A.~Jiménez. This submission reflects the same technical content as the manuscript previously submitted to IISWC~2020 and ISPASS~2021. All research in this submission was performed entirely during the author's time as a graduate student at Texas A\&M University. No part of this work was conducted at, funded by, or related to the author's current employer.}
}

\author{\IEEEauthorblockN{Vikas}
% \IEEEauthorblockA{\textit{Department of ECEN} \\
\textit{Texas A\&M University}\\
% College Station, TX \\
Email - vikas.dce2016@gmail.com\\
\and
\IEEEauthorblockN{Paul Gratz}
\textit{Texas A\&M University}\\
% College Station, TX \\
\and
\IEEEauthorblockN{Daniel A. Jiménez}
\textit{Texas A\&M University}\\
% College Station, TX \\
%\and
%\and
%\and
%\and
}

\maketitle

%ABSTRACT 
\begin{abstract}

  Conditional branch prediction predicts the likely direction of a
  conditional branch instruction to support ILP extraction. Branch
  prediction is a pattern recognition problem that learns mappings
  between a context to the branch outcome.  An accurate predictor
  reduces the number of instructions executed on the wrong path
  resulting in an improvement of performance and energy
  consumption. In this paper, we present a workload characterization
  methodology for branch prediction. We propose two new
  workload-driven branch prediction accuracy identifiers -- branch
  working set size and branch predictability. These parameters are
  highly correlated with misprediction rates of modern branch
  prediction schemes (\emph{e.g.} TAGE and perceptron). We define the
  branch working set of a trace as a group of most frequently
  occurring branch contexts, {\em i.e.}, the 3-part tuple of branch
  address, and associated global and local history. We analyze the
  branch working set's size and predictability on a per-trace basis to
  study its relationship with a modern branch predictor's accuracy. We
  have characterized 2,451 workload traces into seven branch working
  set size and nine predictability categories after analyzing their
  branch behavior. We present further insights into the source of
  prediction accuracy and favored workload categories for modern
  branch predictors.

\end{abstract}

\begin{IEEEkeywords}
branch prediction, working set, predictability
\end{IEEEkeywords}

%SECTION 1 : Introduction 
\section{Introduction}

Branch prediction is a primary contributor to performance in modern
processors. Programs have multiple branch instructions that can change
the direction of the program flow. Branch outcomes are unknown
for many cycles. However, the processor must be continually fed to
exploit the instruction-level parallelism present in a program. Thus,
branch predictors are employed to predict the branch outcome and
improve the flow of instructions through the pipeline. Branch
prediction leads to instructions being speculatively executed on the
predicted path. If the predict is correct, execution continues
uninterrupted. Otherwise, the pipeline must be flushed to get rid of
wrong-path instructions. Based on pipeline depth and instruction
window size, a branch misprediction can lead to large penalties
resulting in wasted time and energy.

An ideal branch predictor would remember the behavior of all previous
branches and the relationships between those branches to the current
branch to give very low misprediction rates. However, a modern branch
predictor with a limited hardware budget does not always have high
accuracy.  Mispredictions can occur due to multiple reasons -- too
many independent branch contexts to remember within the restricted
hardware budget, no recognizable pattern, or the patterns rooted too
deeply in history.  Misprediction rate and the associated penalty
determine the impact on a processor's performance. Much previous work
has precisely quantified this misprediction
penalty~\cite{eyerman2006characterizing}. However, to the best of our
knowledge, there have been very few attempts to analyze the
misprediction rate patterns in workloads and form a branch prediction
based workload-characterization methodology. We note that past
Championship Branch Prediction (CBP) competitions compare the designs
for their performance in application-based workloads groups. This type
of comparison serves the purpose of finding out an overall superior
branch predictor. However, it gives us no further insight into the
relationship between a predictor's accuracy and branch characteristics
of a workload, and whether a design prefers a particular branch
distribution. Applications are evolving at a rapid pace so it is
imperative that we understand a workload's inherent branch behavior
and predictability to estimate the scope of improvement in the current
state of the art in branch predictor designs. Further, a deeper
understanding of the sources of branch misprediction could point to
new directions in branch prediction.

In this work, we show that a branch predictor's accuracy can be judged
by identifying and studying a suitable subset of branch contexts in
the workload, called the branch working set (BWSET). We note that a
tuple of a branch instruction's program counter (PC) and prior branch
history constitutes the most basic unit of branch context, as
used by modern branch predictors.  The branch working set is
defined as a collection of most frequently seen tuples in the
workload.  Our work introduces two new architecture-independent
parameters -- size and inherent predictability of branch working set
to define the branch characteristics of workloads. While the branch
working set size indicates the amount of branch context that needs to
be tracked and learned for prediction, the inherent predictability is
a conservative measure for achievable prediction accuracy. These
parameters strongly correlate with the actual branch misprediction
rate of modern branch predictors. We present results and discussion
for multiple ways to incorporate branch histories in the tuple to
achieve a strong correlation between BWSET size, predictability and
miss-rates for modern branch prediction schemes to present a
meaningful workload characterization. We characterize 2,451 traces
into 7 BWSET size and 9 BWSET predictability bins. The branch
behavior-based categorization provides further insights into a
predictor's accuracy and favoritism for a specific category of
workloads.

This paper makes the following contributions:
\begin{enumerate}

\item We propose the definition of branch working set (BWSET) in
  workloads and identify BWSET's size and predictability as a measure
  for the accuracy of a branch predictor.

\item We empirically show that branch working set's size and
  predictability correlate strongly with the misprediction rates of
  the modern state of the art branch predictors.

\item We present an analysis and characterization framework for a wide
  range of industry provided traces based on its branch working set
  and inherent predictability.

\end{enumerate}

%Needs to be corrected

In this paper, we present the background and related work in
Section~\ref{sec:background}. We review branch prediction
fundamentals and then briefly discuss modern branch prediction schemes
such as TAGE and multiperspective perceptron branch predictor,
along with a discussion about some prior attempts for the
classification of branch behaviors. In Section~\ref{sec:workingset},
we define the branch working set and branch predictability and present
a branch tuple based approach to identify and characterize the branch
behavior in traces. We explain our methodology in
Section~\ref{sec:meth} and results in Section~\ref{sec:results}, and
present a comprehensive analysis of observed trends. We state our
findings and conclude in Section~\ref{sec:conc}.

%SECTION 2 : Background and Prior Work 

\section{Background and Prior Work}
\label{sec:background}

Branch prediction is performed during instruction fetch by identifying
and predicting potential branches in the current fetch group.  The
predictions are later confirmed during execution. An incorrect
prediction leads to a pipeline flush and redirect to the correct path.
Based on the details of the microarchitecture, branch mispredictions
lead to varying levels of performance and energy loss.  Traditional
and modern dynamic branch prediction designs learn branch behavior
autonomously and adapt to changing program behavior.  Branch
prediction was first documented in by Smith~\cite{smith1983branch} and
has been continuously refined since then. The Smith predictor
(proposed by James Smith)~\cite{smith1983branch} is one of the
earliest dynamic branch prediction techniques, where a few bits of
branch PC is used to index a table of 2-bit saturating counters. The
high bit of the counter is the prediction, and the counters are
incremented or decremented as branches are taken or not, respectively.
More complex branch prediction schemes use the current branch PC and a
prior branch history in a two-level organization to give a more
accurate prediction or combine them in to get a single-level
organization.  Yeh {\em et al.}~\cite{yeh1991two, yeh1992alternative,
  yeh1993comparison} note that while the global history branch
predictor is more effective for applications with lots of if-else
constructs, per-address local branch predictors are better for
applications having more loop control instructions. Two-level branch
predictors have higher hardware costs as they require either extended
history tracking or multiple pattern history tables to mitigate
aliasing. To reduce aliasing, McFarling \cite{mcfarling1993combining}
proposed the {\em gshare} and {\em gselect} predictors that
exclusive-or or concatenate the PC and global history.  Due to
exclusive-oring, {\em gshare} can create more uniform indices and thus
reduces aliasing better for a large-sized branch predictor when
compared to {\em gselect}. Some other predictors exploit the branch
bias behavior as well. Lee {\em et al.}~\cite{lee1997bi} propose a
Bi-Mode predictor that partitions the pattern history table (PHT) into
taken/not taken halves called ``direction predictors,'' selected by
the choice predictor. Most of the branches are either biased towards
taken or not taken direction, and the choice predictor remembers this
bias.

\subsection{Modern Branch Prediction schemes}
\begin{figure}[H]
\centering
\includegraphics[width=3.5in]{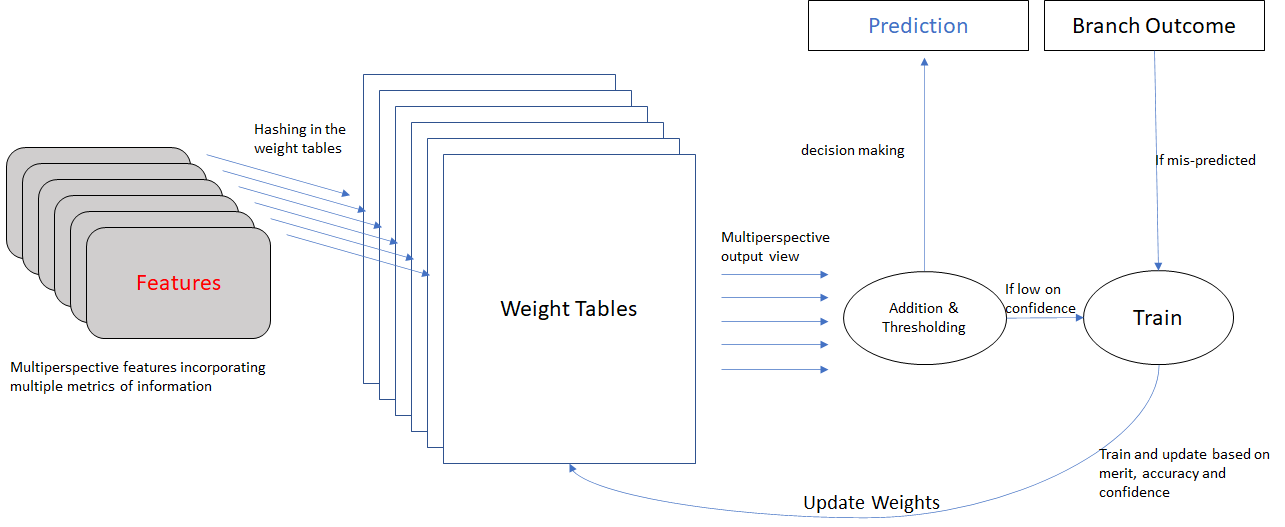}
\caption{Multiperspective perceptron branch predictor}
\label{mppbp}
\end{figure}

Jim{\'e}nez {\em et al.}~\cite{jimenez2001dynamic} introduced a
machine-learning based branch predictor that uses a single perceptron
per branch.  Perceptron information is stored as a vector of multiple
signed weights that show the degree of correlation between various
branches. The perceptron output is the dot product of the branch
history and weights, added with a bias weight. The prediction is taken
if the output is greater than zero, not taken otherwise.  The weight
tables are updated based on the prediction's rightness and confidence
once the branch outcome becomes known. For the same hardware budget,
the perceptron predictor can track and learn branch correlation with a
deeper history than previous predictors. Subsequent work has further
improved the perceptron branch prediction design to improve
latency~\cite{jimenez2003fast}, accuracy for linearly inseparable
functions~\cite{jimenez2005piecewise} and storage budget issues
\cite{loh2005reducing, jimenez2006controlling}. The most recent work
in this domain, multiperspective perceptron branch predictor
\cite{jimenez2016multiperspective1, jimenez2016multiperspective2}, is
essentially a hashed perceptron predictor \cite{jimenez2005piecewise,
  tarjan2005merging} that utilizes a variety of features based on
different organizations of branch histories in addition to the
traditional global path and pattern histories, shown in
Figure~\ref{mppbp}. The authors note that none of the novel features
provide a particularly useful prediction on their own.  However, when
used collectively along with the traditional features, they succeed in
finding alternate perspectives on branch histories and allow finding
new relationships.

While some branches prefer a short history requiring less training
time, others exploit a longer history, achieving higher accuracy by
finding more distant correlations.
Seznec~\cite{seznec2005analysis} proposed a geometric history length
(GEHL) branch predictor in which multiple tables are hashed with
combinations of branch addresses and geometric global/path history
lengths. Using history lengths in geometric series enables the
predictor to track the long history in higher-order tables while still
using most of the tables with shorter
histories. Seznec~\cite{seznec2006case} improves on his GEHL predictor
by introducing the concept of tagged tables in the Tagged geometric
history length predictor (TAGE). This design has a base predictor, a
2-bit bimodal predictor, along with multiple partially-tagged
components that are accessed by geometric history length. Both the
base and partially tagged predictor tables are indexed to give a
branch prediction. If more than one tagged tables get hit, the
prediction is given by the longest history length table; otherwise,
the base prediction is used. The TAGE family of branch predictors
\cite{seznec2014tage, seznec2016tage,seznec201164} has been quite
successful in Championship Branch Prediction competitions (Winner of
CBP5). The recent TAGE predictors \cite{seznec2014tage,
  seznec2016tage} use a perceptron predictor as a ``statistical
correlator'' to find the relationships connecting a branch history and
branch outcome on TAGE algorithm failure.

\begin{figure}[H]
\centering
\includegraphics[width=3.5in]{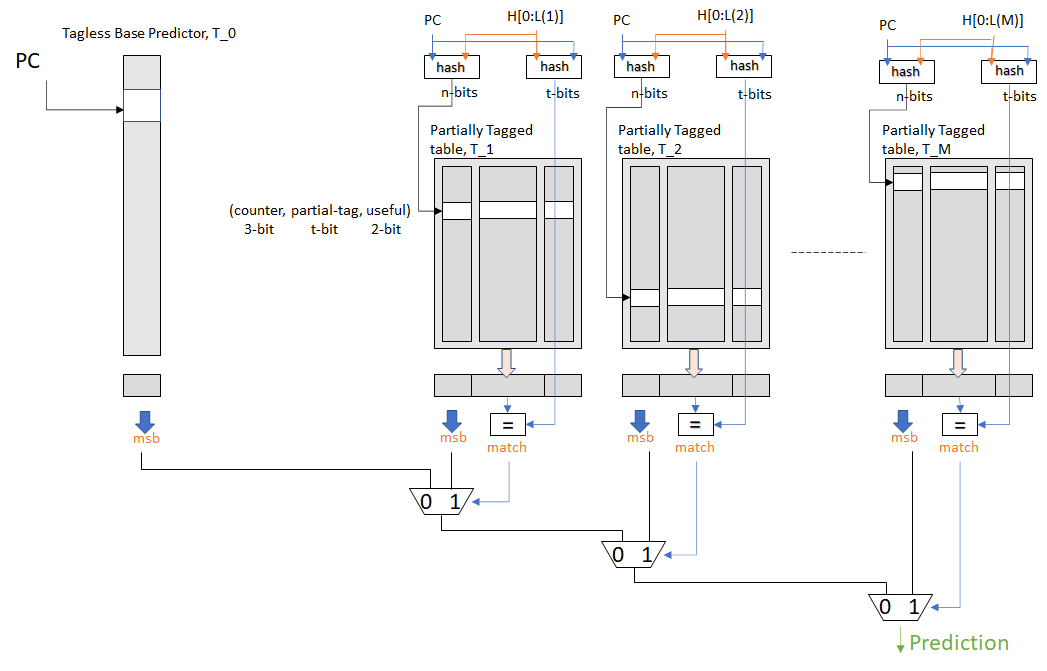}
\caption{TAgged GEometric history length predictor [TAGE]}
\label{tage}
\end{figure}

\subsection{Characterizing branch behavior}
There have been prior attempts to create an architecture-independent
branch characterization to attain a good correlation with prediction
accuracy for various branch predictors. This task is not
straightforward because the prediction accuracy rate not only depends
on the distinct branch characteristics of the workload but also the
predictor's algorithm and physical structure. We saw a wide range of
branch predictors in the literature, each having a branch history
pattern that they can or cannot detect. It is not easy to find a
single characteristic or parameter that relates strongly to prediction
accuracy for a wide variety of modern branch predictors. In the
following sections, we discuss a few such attempts at profiling and
characterizing the branch behavior.

Chang {\em et al.} \cite{chang1996branch} propose a branch
classification mechanism based on taken rate, {\em i.e.}, the fraction
of taken branches out of the total branch occurrence in a program's
execution. A taken rate close to 1 or 0 signifies that branch is easy to
predict as it is biased toward one direction, while a taken rate close
to 0.5 means branch is hard to predict. The disadvantage of this
technique lies in its false classification of some simple to predict
branches as challenging to predict. Haungs {\em et
  al.}~\cite{haungs2000branch} propose a transition rate metric to
alleviate the limitation of the taken rate based method. They define
transition rate as a fraction of branches that change its direction
from all branches in the program's execution. A branch with a low or
high transition rate is said to be easy to predict. A low transition
indicates a branch bias towards one of the outcomes, while a high
transition rate indicates a regular pattern. The author shows that a
branch classification method using both taken and transition rate
parameters show an even superior accuracy. However, this method is
unable to detect a branch with a regular but slightly intricate
pattern.

Chen {\em et al.} \cite{chen1996analysis} show that a simple branch
prediction can be seen as a partial-matching algorithm that uses a
collection of Markov predictors of multiple orders. A ``$n^{th}$''
order Markov predictor uses the most prevalent branch outcome seen in
the last $n$ branches to predict the result of the next branch. Our
tuple definition also contains branch history information of $m$
previous branches, which is used to calculate the predictability
measure for the workload. Chen {\em et al.} use this information to
offer more accurate branch predictors while we use this information to
classify the branches.

Yokota {\em et al.}~\cite{yokota2008potentials} propose a branch
classification based on the information theory concept of branch
entropy. This work calculates the entropy of the branch outcome using
conventional entropy
formula,\\

$E = - \sum{\log_2{p(S\_i)}. p(S\_i)}$

$S\_i$ denotes all potential branch result patterns and $p(S\_i)$ the
probability of branch pattern $S\_i$. This work notes that entropy
correlates with the misprediction rate of a branch
predictor. Additionally, this work also estimates the topmost bound
for the accuracy of a predictor.  This method is complex to implement,
and a simplified version of this method is performed by Pestel {\em et
  al.}~\cite{de2016linear}. They introduce linear branch entropy as a
parameter to measure the regularity in the branch behavior of a
workload. For each set of PC and branch history, they calculate the
probability of the taken outcome. Their taken rate definition is
different from that of Chang {\em et al.}~\cite{chang1996branch} as
they figure taken probability for every combination of history pattern
and branch PC. They estimate linear branch entropy using the following
formula, where $p$ is the
probability of a taken branch, given specific branch history. \\

$E\_L(p) = 2. min (p, 1-p)$

As per this equation when $p=1$ or $p=0$ linear branch entropy equals
to 0 (biased branch, highly predictable), and it equals 1 (random
branch behavior, lower predictability) when $p=0.5$. Linear branch
entropy is averaged for the set of branch PC and its history to get a
single linear branch entropy number for the trace. They empirically
explain that linear branch entropy relates better with branch
predictor accuracy when compared to work done
in~\cite{yokota2008potentials}. They use this branch profiling
information to create a model that relates branch entropy with the
predictor's misprediction rate. They use this relation to predict the
misprediction rates for different branch prediction strategies based
on their model for design space exploration.

Our work also identifies the branch and its associated history as the most
basic unit for characterizing a branch's behavior. Pester {\em et
al.}~\cite{de2016linear} directly use this information to
create their linear branch entropy model, while this paper introduces an
additional level of profiling for characterization where we identify a sub-set
of total branch contexts present in a workload based on occurrence count. We
do this exercise to determine the most frequent contexts.  We call this
sub-set a branch working set and then relate the size of the branch and its
predictability with a modern predictor's accuracy. This paper presents a
robust yet straightforward characterization methodology validated with the
help of a much broader set of industry-provided traces from various
application domains.

%SECTION 3 : Branch Working Set and Predictability for Workloads 
\section{Branch Working Set and Predictability for Workloads}
\label{sec:workingset}

% It has also been established that branch context lies not just in the
% PC value, but both branch PC and histories together constitute the
% actual branch context. - by who, needs citation and rewrite

In Section~\ref{sec:background}, we noted
that initial branch prediction ideas revolved around remembering the
outcome of all previous branches (global branch history), each of the
individual branches uniquely identified by its branch PC (local branch
history), or both. In this work, we propose a technique to
characterize branch behavior independent of both the application
category of the trace or the branch predictor that is used.  We aim to
characterize workloads with respect to their branch characteristics
and ease of prediction based on the concepts of bias and aliasing. We
note that in any practically sized predictor, aliasing is bound to
happen. Thus, having a large number of basic branch context units (PC
or PC+history) in any trace leads to increased aliasing and high
misprediction rates. Additionally, having more bias for a branch
context means having higher inherent predictability in the trace.

%3.1 Branch Working Set
\subsection{Branch Working Set (BWSET)}
A branch context distinguishes any predictor's behavior for a specific
branch.  We define branch working set as the subset of branch
contexts, which determines the most frequent and prominent branch
contexts present in the trace based on their occurrence count and a
threshold, $\theta$. We keep this threshold value high enough to
capture the prominent branch context and disregard some stray or
infrequent branch contexts. The branch working set size indicates the
extent of all the context which needs to be tracked and learned from
the workload. We show that the branch working set size correlates with
the misprediction rate of a predictor because more branch context
means more aliasing in any practically sized modern predictor. Thus,
the branch working set can be a straightforward yet beneficial measure
for workload characterization based on ease of branch prediction. This
measure is independent of predictor architecture and is determined
based on branch behavior present in the workload.

%3.1.1 Tuple based Branch Working Set
\subsubsection{Tuple based branch working set}
\begin{figure}[H]
\centering
\includegraphics[width=3.5in]{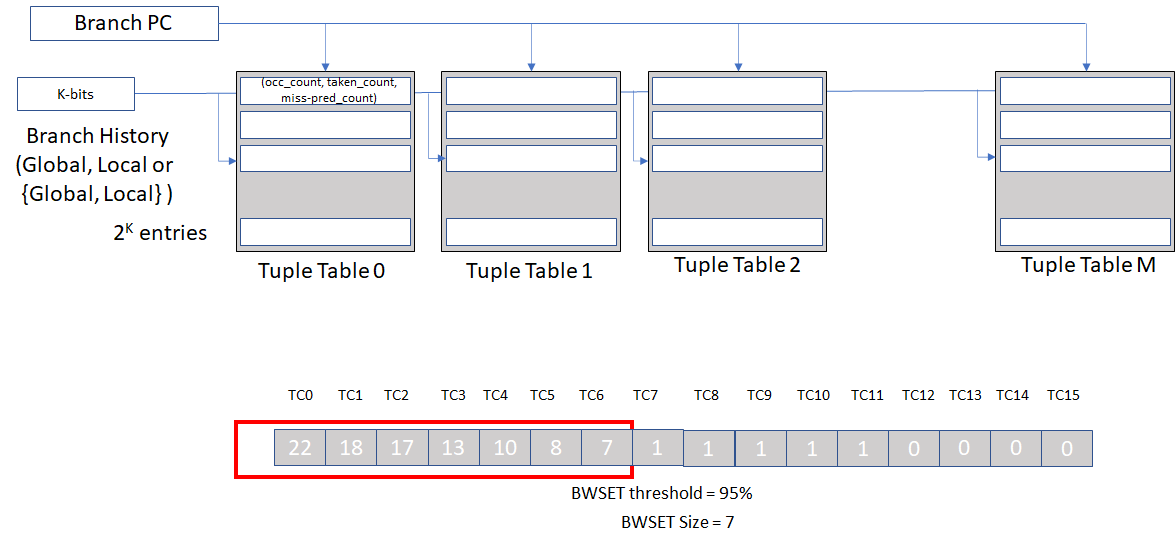}
\caption{Tuple based working set definition}
\label{tuplebasedbwset}
\end{figure}

For a first-level analysis, we assume that the branch context contains
only the most fundamental branch identifier, the PC. But a strong
correlation between the history of prior branches with the result of
the current branch ensures a low misprediction rate in predictors. The
confidence of a branch outcome for a particular history pattern
determines the accuracy of any branch prediction scheme. We employ
this knowledge to update the branch working set definition to
incorporate the branch history with the conditional branch PC.  As
seen in branch prediction literature, local histories are rarely used
alone; they are usually combined with a section of global histories in
multiple different setups to access the prediction tables for high
accuracy branch prediction. We define the basic unit of branch context
in the form of a ``global tuple'' and ``global-local tuple''
consisting of branch PC and few bits of global and local branch
history preceding the current branch. The proposed tuple based branch
profiling data structure is shown in Figure~\ref{tuplebasedbwset}. We
maintain one table per PC and increment the count of a previously seen
tuple or allocate a new entry in the table when we see a new
tuple. Each table entry contains the occurrence count, taken count,
and misprediction count for each of the tuples.

This methodology gives us a way of combining a branch PC and different
global and local branch history scenarios under which the prediction
is made. Here, the number of elements in a branch working set would be
the number of unique tuples, comprising 95\% ($\theta = 95\%$) of
total tuples seen in the entire trace. Figure \ref{tuplebasedbwset}
shows that only seven tuples constitute more than 95\% of total
branch/tuple occurrence count. This analysis helps us categorize the
trace pool we have based on this definition of branch working set. We
run this study for global history lengths, N=8, 16, 24, 32, 48 and 64 and
local history lengths, M=4, 8, 16 and 24 with $\theta = 95\%$ to
determine the optimum branch history lengths in each case to see the
correlation between branch working set size and misprediction rates of
modern predictors. We limit our studies' scope to global and local
histories because they are a compelling yet straightforward way
of tracking branch histories and is utilized in most of the branch
predictor designs.

%3.2 Predictability in the Branch Working Set
\subsubsection{Predictability in the Branch Working Set}
We define predictability based on the inherent bias of a branch
context, branch PC, Tuple, or Triplet and the calculation formula is
as below:

Predictability = $\frac{max(taken count, not-taken count)}{occurrence\_count\_of\_branch\_context}$

We average this predictability parameter for all unique branch
contexts present in the branch working set to get one quantitative
predictability measure for the entire trace. A predictability value
close to 100\% implies that branch context is entirely biased towards
either taken or not taken, and the workload is bound to show a very
low misprediction rate in any practical design. In our framework, the
predictability value is bounded by a lower minimum of 50\%. This value
indicates that branch behavior is very random and difficult to track,
leading to high misprediction rates.

%SECTION 4 : Methodology 

\section{Methodology}
\label{sec:meth}

%4.1 Workloads
\subsection{Workloads}
A key component of any such study is a broad set of workloads to
achieve high confidence correlations for meaningful observations.  To
this end we leveraged the traces released for the fifth Championship
on Branch Prediction (CBP5) in 2016~\cite{cbp5} and for the
Championship Value Prediction competition (CVP'18)~\cite{cvp}. 
CBP5 provided 440+ traces generated by Samsung.  These traces
have 4 categories – long mobile, short mobile, long server, and short
server. 
CVP provided an additional 2,014 publicly released traces,
generated by Qualcomm, in 4 application domain categories –
compute-int, compute-fp, server, and crypto workloads.
Despite the traces being released for a different purpose, we find
they accurately represent real workload branch behavior and the trace
format contains sufficient detail to be used in this study by our
branch profiling and analysis framework.  Thus, we examine a total of
2,451 industry released traces, given in Table~\ref{workloads}, for
defining characterization parameters – branch working set and
predictability.  By studying the trends and correlations of these
parameters with misprediction rates of two of the most prominent
modern branch predictors, we characterize these traces in several new
categories beyond the boundaries of any traditional application
domain-dependent characterization.

\begin{table}[h!]
\centering
\caption{Workload characterization based on application group and source}
\begin{tabular}{|l|l|l|l|}
\hline
\textbf{Application-group} & \textbf{Source} & \textbf{Released for} & \textbf{No. of traces} \\ \hline
$LONG\_SERVER$ & Samsung & CBP5 & 8 \\ \hline
$LONG\_MOBILE$ & Samsung & CBP5 & 31 \\ \hline
$SHORT\_SERVER$ & Samsung & CBP5 & 293 \\ \hline
$SHORT\_MOBILE$ & Samsung & CBP5 & 105 \\ \hline
$COMPUTE\_INT$ & Qualcomm & CVP & 982 \\ \hline
$COMPUTE\_FP$ & Qualcomm & CVP & 140 \\ \hline
$SRV$ & Qualcomm & CVP & 786 \\ \hline
$CRYPTO$ & Qualcomm & CVP & 106 \\ \hline
\end{tabular}
\label{workloads}
\end{table}

%4.2 Analysis Framework
\subsection{Analysis Framework}
\begin{figure}[H]
\centering
\includegraphics[width=3.5in]{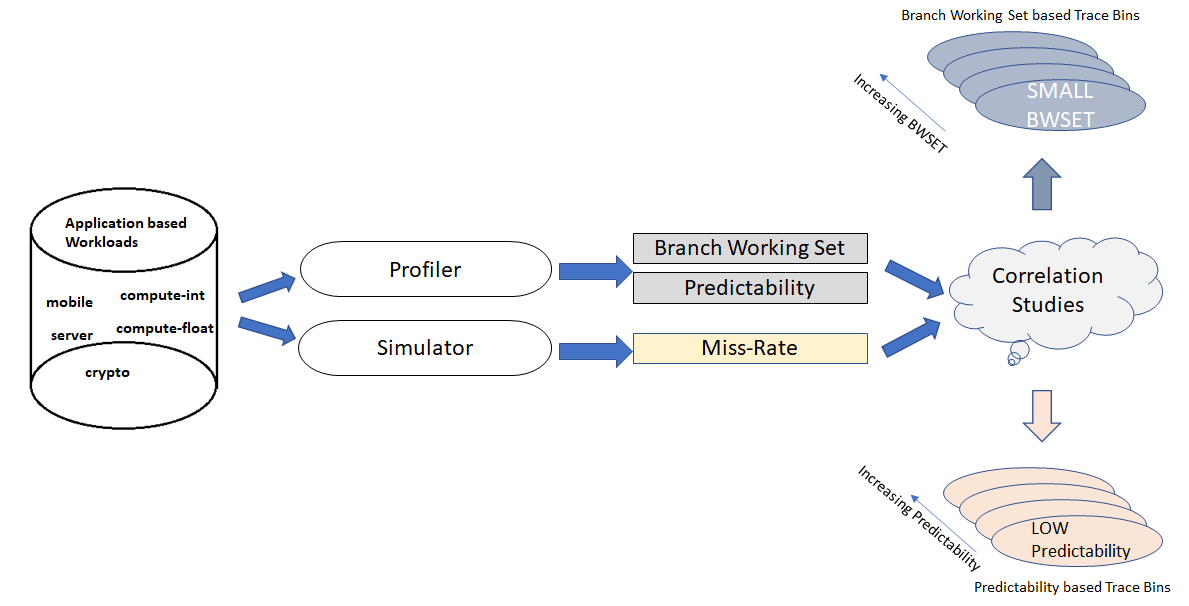}
\caption{Workload characterization and analysis framework - based on branch working set and predictability}
\label{framework}
\end{figure}

We have extended the CBP5's infrastructure to create our branch
profiling and analysis framework, shown in
Figure~\ref{framework}. This framework provides the data to
characterize a given trace with respect to branch working set's size
and predictability. The motivation behind using this existing
infrastructure was to access the trace reader and simulation set-up
provided for CBP5. The CBP5 infrastructure core offers a way to
simulate state-of-art branch prediction designs to correlate with a trace's
extracted branch characteristics. We make the following enhancements
to the CBP5's existing infrastructure to create our branch profiler
and analysis framework.

%Trace Reader for CVP
\textbf{Trace Reader for CVP'18 traces} - In our framework we extended
the CPB5 infrastructure to read CVP traces along with CBP5 traces.
% This is a lot of detail that noone will care about for the IISWC
% paper: We first need to uncompress CVP'18 traces and then identify
% all the branch instructions based on the given trace format. We
% extract branch characteristics for conditional branch instructions
% only and then use this data in our analysis framework. Within this
% modified CBP infrastructure, an appropriate trace reader is called
% upon based on whether we run Samsung's BT9 trace format or
% Qualcomm's compressed trace format.

%Branch Data Logging
\textbf{Branch Data Logging} - The framework has provisions for
simulating a branch prediction scheme and logging multiple useful
information on the fly.  The framework can be configured for a PC
only, Global Tuple ({PC, N-bit global history}), and Global-Local
Tuple ({PC, N-bit global history, M-bit local history}) based data
logging. In PC mode, a flag indicates whether a branch is static or
dynamic. For Tuple modes, the framework keeps track of all the global
and local branch histories and associates the most recent bits of
branch history with the current branch PC to create a branch
tuple. The occurrence count, misprediction count, and taken count for
all the unique tuples (pair of history and branch PC) in the workload
trace is logged as well. Due to the involvement of history and the
branch PC, each of the tuples is uniquely identifiable. Even the same
branch PC with a regular alternate taken and the not-taken pattern is
categorized in two different tuples. An amalgamation of global and
local histories and branch PC results in a close tracking of a modern
predictor's performance.

%Post-processing
\textbf{Post-processing to define branch working set size and
  predictability}.  Post-processing for the logged branch data is done
after all the branch instructions have been simulated. The first step
is to identify the branch working set for the trace. All the branch
data-structure (PC only, or Tuple) is reordered based on the
decreasing order of its occurrence count. We then select all the most
highly occurring branches in the branch working set such that their
cumulative occurrence count is more than the total branch occurrence
count in the trace. After defining the branch working set, we take a
weighted average of the predictability numbers for all set elements to
generate one representative predictability measure for the given trace. 
We calculate the branch misprediction rate
in the form of mispredictions per kilo branches (MPKB)
for all the elements present in the based branch working set. The
MPKB numbers for all 2451 traces are studied for their relationship
with the respective branch working set's size and predictability
numbers to identify trends for workload characterization.

%4.3 Characterization based on BWSET's size
\subsection{Characterization based on branch working set's size}
A workload trace having 
a large working set size indicates that there
is much more context that needs to be tracked and learned for
effective branch prediction. In a practically sized predictor, this
leads to increased destructive interference and higher misprediction
rates. Similarly, a small branch working set indicates that branch
contexts can be easily learned by any modern generic predictor,
leading to a very low misprediction rate. Thus, we expect branch
working set size to be a good branch behavior indicator for
prediction ease in a given trace. We undertook studies to create a
branch profile of traces for multiple configurations by varying branch
working set threshold $\theta$ and the amount of global and local
history in the tuples and triplet, $N$ and $M$ respectively. We
propose a characterization methodology where we categorize traces in
multiple branch working set bins. We categorize traces based on whether they have small, medium, or
large branch working set sizes. We further subdivide into seven
subcategories, as given in Table~\ref{tupletrace2}.  
We decided on having buckets with exponentially
increasing size to cover the entire range.We study BWSET size vs. MPKB 
in multiple BWSET containers to understand the
relationship. The aim is to find a framework configuration to
characterize workload traces for their ease of prediction
successfully.

\begin{table}[h!]
\centering
\caption{BWSET Size bins for workload characterization}
\begin{tabular}{|l|l|}
\hline
\textbf{BWSET Size} & \textbf{Characterized BWSET bin} \\ \hline
1 - 100 & BWSET-LOW1 \\ \hline
100 - 1k & BWSET-LOW2 \\ \hline
1k - 10k & BWSET-MEDIUM1 \\ \hline
10k - 100k & BWSET-MEDIUM2 \\ \hline
100k - 1M & BWSET-HIGH1 \\ \hline
1M - 10M & BWSET-HIGH2 \\ \hline
$>$10M & BWSET-HIGH3 \\ \hline
\end{tabular}
\label{tupletrace2}
\end{table}

%4.4 Characterization based on predictability in branch working set
\subsection{Characterization based on predictability in branch working
  set} 
Branch working set alone (however a tuple is defined) is
insufficient to accurately characterize the misprediction rate of the
workload.  This is because if the outcome of branches is random, even
the smallest working set will not be predictable in any realistic
predictor.  Thus, we endeavor to quantify the predictablity of the
tuples in the workload.  In the proposed framework, we calculate a
quantitative predictability measure for the workload trace based on
the inherent bias of the tuple based branch contexts. We estimate this
predictability measure by averaging individual predictability values
on a per-tuple basis in our defined branch working set.  We will
examine how well predictability correlates with any reasonably
accurate branch prediction scheme and whether it can be a metric in
itself to characterize workloads for branch prediction in the
following sections. We define several predictability
bins as shown in Table ~\ref{predtrace2} and study the relationship between predictability and
misprediction rate for state-of-art branch prediction schemes. 

\begin{table}[h!]
\centering
\caption{BWSET Predictability bins for workload characterization}
\begin{tabular}{|l|l|}
\hline
\textbf{Predictability in BWSET} & \textbf{Characterized Predictability bin} \\ \hline
$<75\%$ & Pred-VLOW1 \\ \hline
75\% - 80\% & Pred-LOW1 \\ \hline
80\% - 85\% & Pred-LOW2 \\ \hline
85\% - 90\% & Pred-LOW3 \\ \hline
90\% - 92.5\% & Pred-MEDIUM1 \\ \hline
92.5\% - 95\% & Pred-MEDIUM2 \\ \hline
95\% - 97.5\% & Pred-HIGH1 \\ \hline
97.5\% - 99\% & Pred-HIGH2 \\ \hline
99\% - 100\% & Pred-HIGH3 \\ \hline
\end{tabular}
\label{predtrace2}
\end{table}

%\pg{Be careful in your writing that you are not making statements
%  before you prove them.  At this stage in the paper you haven't shown
%  any results but there is lots of ``This predictabilty metric
%  correlates with blah blah'' type language.  This is jumping the gun,
%  you haven't shown that yet (that's the point of the results
%  section).  I addressed some of it but reread and fix were I did not.
%  Also, nobody cares about the tools work, just say what your tool can
%  do, not how it does it (unless that is somehow novel, but reading
%  traces is straight foward).  Finally, this text talks about both
%  TAGE and perceptron, if we don't have perceptron results then you
%  need to get that out of here too (though maybe we should add it back
%  as a final section if we have space)}

%SECTION 5 : Results and Analysis
\section{Results and Analysis}
\label{sec:results}
In this section, we discuss the workload characterization results for
all the configuration modes based on branch PC, global tuple, and
global-local tuple.  We identify the new workload characterization
groups that we have defined depending on both branch working set size
and predictability. We analyze our results to study the relationship
of the extracted branch behavior parameters with Mispredictions Per
1000 Branches (MPKB) of state-of-the-art branch predictors for
different analysis modes and the length of the history. 
The branch working set size and predictability are design-independent parameters 
and are expected to correlate well with the misprediction rate for any
advanced predictor, whether TAGE or perceptron. Thus, going forward we present our discussion based on TAGE numbers only. 
Later, we briefly discuss results for the multiperspective perceptron
predictor to see if these trends hold true across these two state-of-art branch prediction schemes. 

%5.1 BWSET based on branch PCs

\subsection{BWSET based on branch PCs}
We first characterize traces based on branch PC as a first-order
analysis. By setting the threshold factor of 95\%, the framework
becomes more exclusive and filters out more low-frequency branch PCs
which have less impact on performance. The branch working set
definition makes more sense for dynamic branches as they change their
direction during a program's execution, thus making it difficult to
track and predict than a static branch that does not alter its course.

\begin{figure}[H]
\centering
\includegraphics[width=3in]{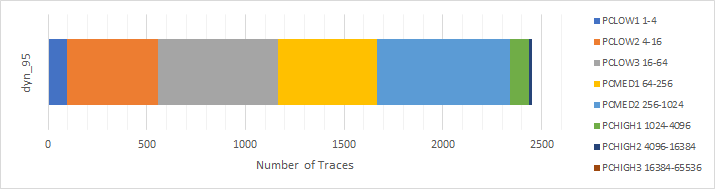}
\caption{Trace distribution based on BWSET of branch PCs}
\label{pctracedistri}
\end{figure}

\begin{figure}[H]
\centering
\includegraphics[width=3.5in]{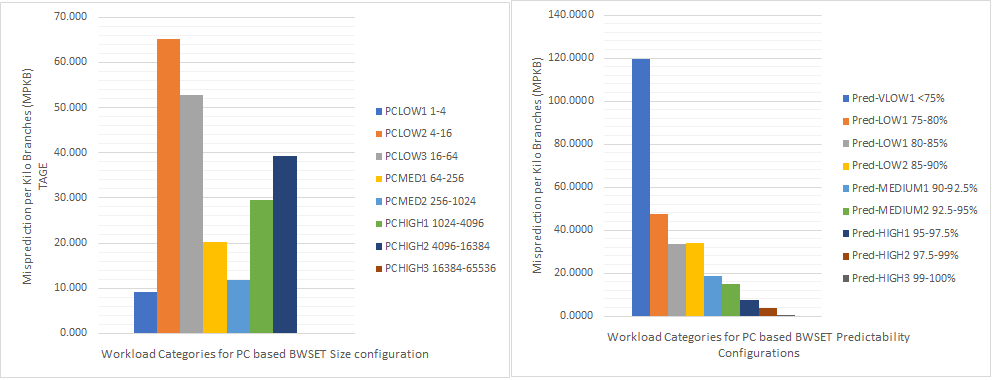}
\caption{TAGE : MPKB vs BWSET Size and Predictability in branch PC based working set }
\label{PCorrelationAS}
\end{figure}

We divide the PC based branch working set range into multiple BWSET size bins
and show the distribution in Figure ~\ref{pctracedistri}.  We average the
MPKB numbers for traces in each bin resulting in a representative MPKB
number as the miss-rate identifier for the trace category. We show the
relationship between branch working set categories and TAGE's misprediction
rate in left part of Figure~\ref{PCorrelationAS}. The presented PC based BWSET
charaterization results in no identifiable correlation hinting that PC alone
does not define the entirety of branch context used by a modern branch
predictor, such as TAGE.

We show the relationship between TAGE MPKB and PC based BWSET predictability
in right part of Figure~\ref{PCorrelationAS}. Even though predictability does
show an inversely proportional relationship with misprediction rate, the
shortcomings of PC based characterization is evident from Table
~\ref{predcomparisonPC}.  In this characterization methodology, most of the
traces have been characterized in low predictability bins and if we compare
the predictability projection vs actual TAGE accuracy, TAGE performs much
better than the projected numbers. The low predictabilty is indicative of the
fact that this is a poor metric and that modern branch predictors do better
than that predictability, thus PC alone is insufficient

%\pg{This needs to be correlated back real branch predictors to show
%  that this is a poor metric, leading us to the tuple metrics.  Add
%  para here of analysis.  ie. }

\begin{table}[h!]
\centering
\caption{Predictability Projection vs TAGE Accuracy for PC based Characterization}
\begin{tabular}{|p{1.50cm}|p{1.75cm}|p{0.75cm}|p{1.50cm}|p{1.35cm}|}
\hline
\textbf{Category} & \textbf{Predictability Range} & \textbf{No of Traces} & \textbf{Predictability Projection} & \textbf{TAGE Accuracy Percentage} \\ \hline
Pred VLOW1 & $ 75\% - 80\%$ & 99 & $70.33\%$ & $88.06\%$ \\ \hline
Pred LOW1 & $ 75\% - 80\%$ & 378 & $78.16\%$ & $95.24\%$  \\ \hline
Pred LOW2 & $ 80\% - 85\%$ & 691 & $83.21\%$ & $96.63\%$  \\ \hline
Pred LOW3 & $ 85\% - 90\%$ & 752 & $87.06\%$ & $96.59\%$  \\ \hline
Pred MEDIUM1 & $ 90\% - 92.5\%$ & 172 & $91.28\%$ & $98.12\%$  \\ \hline
Pred MEDIUM2 & $ 92.5\% - 95\%$ & 97 & $93.65\%$ & $98.50\%$  \\ \hline
Pred HIGH1 & $ 95\% - 97.5\%$ & 90 & $96.20\%$ & $99.23\%$  \\ \hline
Pred HIGH2 & $ 97.5\% - 99\%$ & 69 & $98.29\%$ & $99.63\%$  \\ \hline
Pred HIGH3 & $ 99\% - 100\%$ & 103 & $99.72\%$ & $99.93\%$  \\ \hline
\end{tabular}
\label{predcomparisonPC}
\end{table}

%5.2 BWSET based on branch Global Tuples 
\subsection{BWSET based on global branch tuples}
We have seen in Figure~\ref{PCorrelationAS} that branch PC is just one part of
the branch context and does not entirely define the branch behavior even for a
very simplistic characterization model. It is crucial to consider both
conditional branch PC and prior branch histories for setting the branch
context. Using global history gives us a simplistic enough model, which still
correlates very well with the modern branch predictor's misprediction rate. We
present the results for a global tuple based independent characterization
model in this section.

%5.2.1 Trace Categorization based on BWSET
\subsubsection{Trace categorization based on branch working set}
\begin{figure}[H]
\centering
\includegraphics[width=3.5in]{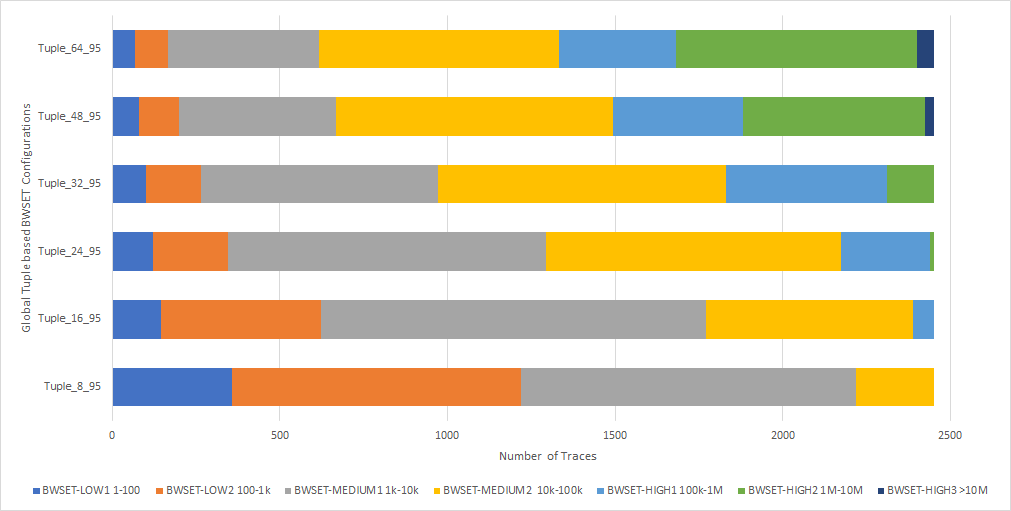}
\caption{Trace distribution based on BWSET of Global Tuples}
\label{tupletracedistri}
\end{figure}

In the tuple based analysis methodology, we identify more unique branch
contexts for each branch PC. We associate a certain length of global history
with the branch PC to give a simple yet powerful measure for branch context in
the trace. Due to these additional criteria, we have many more identifiable
branch contexts than the previous methodology for just the PC based branch
working set. We have tried several different history lengths ranging from 8 to
64 with BWSET $\theta = 95\%$. With increasing history bits in the tuple, the
range of branch working set sizes shown by the traces also increases. But this
increase for any particular trace is not exponential, as all possible history
values are not seen when we increase the amount of history information in our
tuple. We note that with increasing branch history, many more traces shift from
low branch working set categories to high branch working set groups. All the
workloads have been characterized in the discussed tuple based BWSET bins, and
the distribution is shown in Figure~\ref{tupletracedistri}.

%5.2.2 Miss Rate vs. BWSET Size
\subsubsection{Miss Rate vs. Branch Working Set Size}
\begin{figure}[H]
\centering
\includegraphics[width=3.5in]{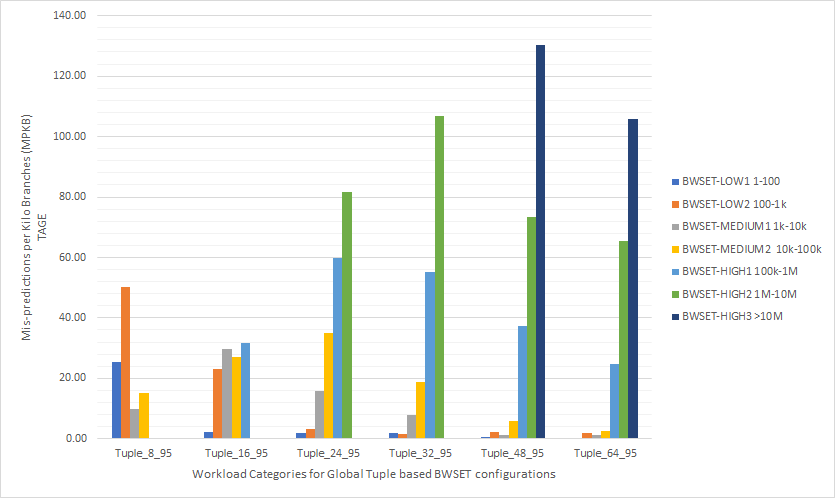}
\caption{TAGE: MPKB vs. BWSET based on Global Tuples }
\label{TuplecorrelationAS}
\end{figure}

In this section, we analyze the correlation of branch working set size with
the representative misprediction per 1000 branches (MPKB) number for TAGE. We
expect to see a high correlation between a predictor's accuracy and the
amount of branch context it tracks, learns, and uses to predict the branch
outcome. All the practical branch predictors are hardware constrained. With
the increasing amount of branch context, these predictors tend to lose
prediction accuracy due to destructive interference in limited sized branch
prediction tables. Again, we note that most of the traces having similar
branch working set sizes show a similar miss rate numbers for both TAGE and
Perceptron predictors. Figure~\ref{TuplecorrelationAS} shows the relationship
between TAGE~\cite{seznec2016tage} predictor's accuracy with BWSET size in all
6 configuration modes. 
%We have measured a predictor's accuracy in terms of MPKB.

A workload trace with a larger branch working set size tends to show a
lower accuracy than traces with a smaller branch working set. We note
that a high correlation is emerging between MPKB and branch working
set size for tuple definition of PC and more than 24 bits of global
history outcome. For history lengths lower than 24 bit, we do not
observe an identifiable pattern. A tuple definition with smaller
global history is unable to uniquely determine all the recognizable
branch context in a trace, thus showing not much correlation with the
predictor's miss rate. The correlation pattern for tuple definition
with history length, N = 32, 48 and 64, is also reasonably regular but
with anomalies and a much higher number of outliers. A tuple with a
large amount of history is also unable to give good correlation
results because the branch contexts get lost in longer histories for
some specific traces, thus showing some deviation from the expected
correlation results. We shall see much more evident trends for a
higher amount of history in tuples when we discuss predictability
based bins in the later section.

\begin{figure}[H]
\centering
\includegraphics[width=3.5in]{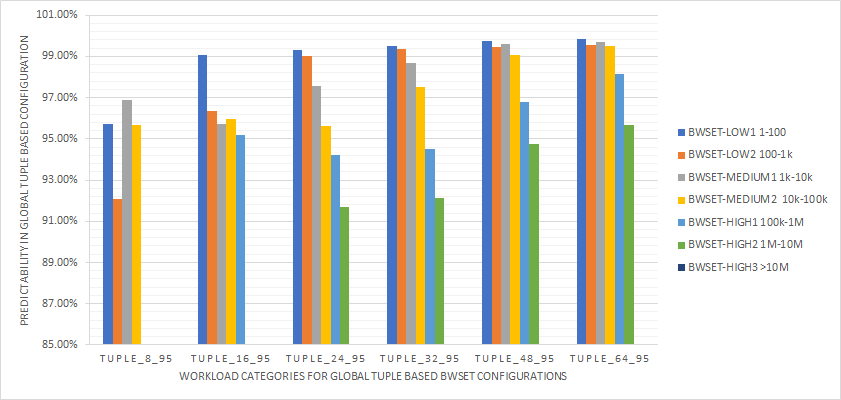}
\caption{Predictability in Global tuple based working set }
\label{predPTuple}
\end{figure}

%5.2.3 Predictability in BWSET bins
\subsubsection{Predicability in Branch Working Set Bins}
In Figure~\ref{predPTuple}, we show the predictability results for each of the
newly defined trace categories. These results justify the trends in miss rate
vs. working set size we observed in the last section. These predictability
numbers are based on the inherent bias in the tuple. We note that a very high
predictability number, {\em i.e.}, at least 98\%, leads to high branch
prediction accuracy. Looking at these plots, we can see why the history length
of 24 shows such a regular correlation pattern. It also indicates a high
predictability correlation in the branch working set with the miss rates of
modern predictors. With branch histories smaller than 24, the tuples do not
show a regular predictability plot as history is too short to fully identify
the majority of the branch context in trace sets that we use. The longer
histories tend to have noisy behavior, which leads to useful context getting
lost in multiple individual low-frequency tuples (filtered out in branch
working set definition).

%5.2.4 Trace Categorizations based on branch predictability
\subsubsection{Trace categorizations based on branch predictability}
\begin{figure}[H]
\centering
\includegraphics[width=3.5in]{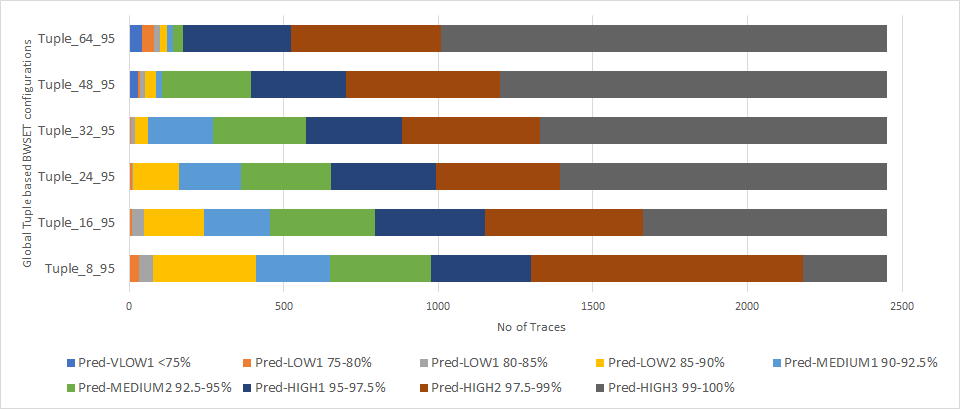}
\caption{Trace distribution based on Predictability in Global Tuple BWSET}
\label{tupletracedistri2}
\end{figure}

We note that predictability measure for a trace has a substantial correlation
with a predictor's accuracy. The predictability data show a trend of their
own. Due to the way the predictability parameter is defined, it gives the
probability for a tuple's biased direction. We average this predictability
characteristics across all the tuples in the branch working set to get a
branch prediction easiness indicator. We see sense in defining the
predictability parameter for only the branch working set because we do not
want predictability to be affected by the behavior of low-frequency tuples.
Figure~\ref{tupletracedistri2} shows predictability based trace distribution
for all the six configuration modes. With increasing predictability, we see
some lower predictability traces moving into high predictability trace
categories until N=32. For longer history lengths, we see several traces losing
their earlier high predictability status and proceed to low predictability
categories.

\begin{figure}[H]
\centering
\includegraphics[width=3.5in]{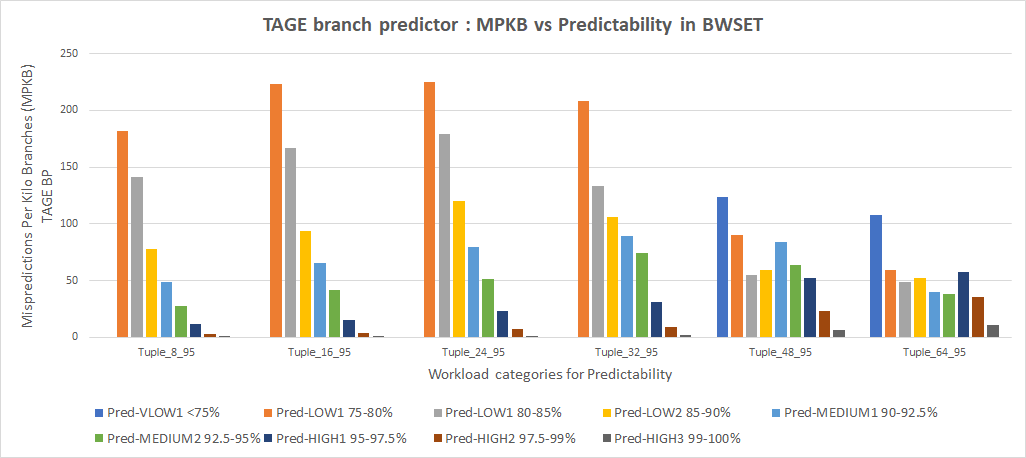}
\caption{TAGE: MPKB vs. Predictability in Global Tuple BWSET }
\label{predcorrAS}
\end{figure}

\textbf{Predictability vs. Miss Rate}: We run the correlation studies
for these predictability based trace characterization in each of the 6
tuple based configuration modes. Figure \ref{predcorrAS} shows the
relationship between TAGE~\cite{seznec2016tage} predictor's accuracy
with the inherent predictability present in BWSET.

In the figure, we see a strong correlation emerging for even smaller
history lengths, N = 8, 16, 24, and 32. There is a fairly regular
pattern that indicates that a trace having higher overall
predictability also has a higher prediction accuracy for the state of
the art branch predictor– TAGE \cite{seznec2016tage}. In the prior
trends seen for BWSET size vs. MPKB in Figure~\ref{predcorrAS}, we
have already established how tuples with N=8 and 16 are not able to
determine all the identifiable branch context in the workload
uniquely. This shortcoming resulted in no visible correlation between
the modern predictor's accuracy with BWSET size. Similarly, seeing the
trends for N=48 and 64, we establish with large amounts of history
that branch context is lost due to aliasing in a hardware limited
modern predictor, resulting in an unclear correlation between MPKB and
inherent predictability BWSET. This observation concludes that a tuple
consisting of PC and 24 bits of global history presents a sweet spot
in workload characterization for branch prediction accuracy. This
configuration shows a strong correlation between MKBP and BWSET size
as well as BWSET predictability.

The accuracy of TAGE is better than the inherent predictability numbers we
project for workloads in medium and high predictability workload bins while it
is marginally lower in low predictability bins , as is shown in
Table~\ref{predcomparisonGlobal}.  The predictability measure reported here is
generally conservative because we work with only generalized global histories
in this part of our study.

%5.4 BWSET based on Global-Local Triplets 
\subsection{BWSET based on global-local branch tuples}
\begin{figure}[H]
\centering
\includegraphics[width=3.5in]{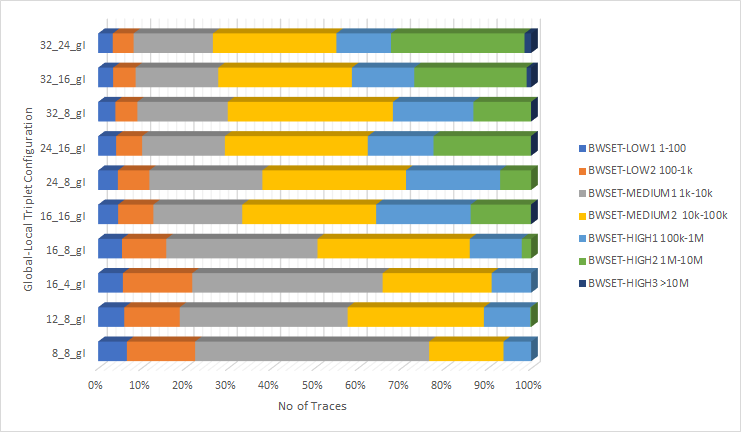}
\caption{Trace distribution based on BWSET size for Global-Local Triplets}
\label{gltupletracedistri}
\end{figure}

\begin{figure}[H]
\centering
\includegraphics[width=3.5in]{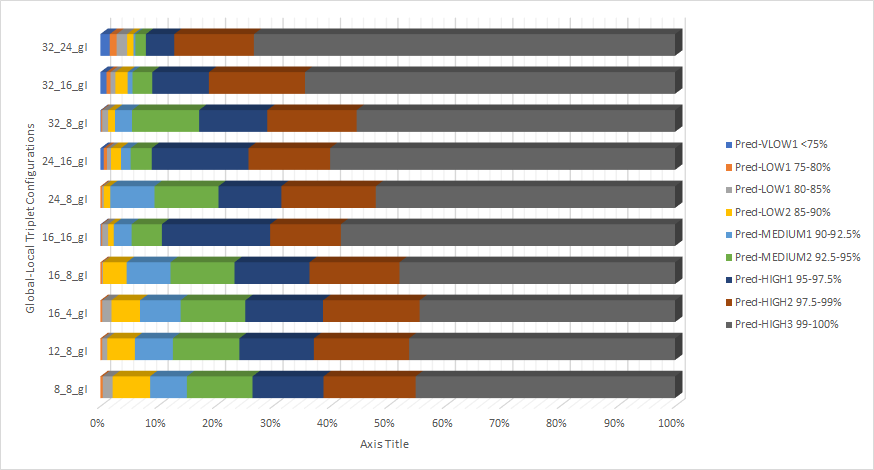}
\caption{Trace distribution based on BWSET predictability for Global-Local Triplets}
\label{glpredtracedistri}
\end{figure}

In the previous section, we have seen that 24 bits of global history
lead to a successful workload characterization. The other components
of branch history, such as local history or staggered histories, are
comparatively challenging to track and increase the complexity of the
characterization model. They need to be monitored and used in addition
to the global histories. But, most modern branch predictors have
certain techniques to use to achieve high prediction accuracy. Thus,
to make our analysis more consistent with modern designs, we present a
discussion on workload characterization based on a 3-way tuple of PC,
global, and local branch history. We use different configurations
where we use 8 to 32 bits of global history, 4 to 24 bits of local
history, and the branch PC to define the global-local tuple. The trace
distribution for configurations based on BWSET size and BWSET
Predictability is shown in Figures \ref{gltupletracedistri} and
\ref{glpredtracedistri}, respectively.

\begin{figure}[H]
\centering
\includegraphics[width=3.5in]{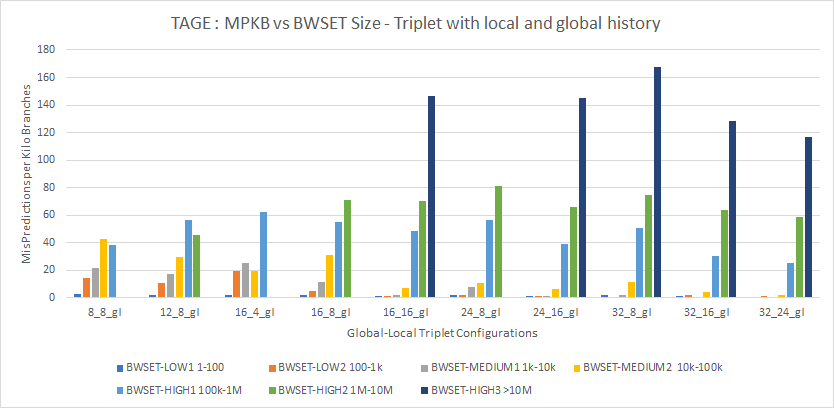}
\caption{TAGE: MPKB vs. BWSET based on Global-Local Triplet }
\label{GLTuplecorrelationAS}
\end{figure}

We compare MPKB vs. BWSET size for the TAGE branch predictor, shown in
Figure \ref{GLTuplecorrelationAS}. When combined, we find that 16 or
more bits of global history along with 8 or more bits of local history
in the tuple definition give a reliable correlation where increasing
BWSET size leads to decreasing prediction accuracy.

\begin{figure}[H]
\centering
\includegraphics[width=3.5in]{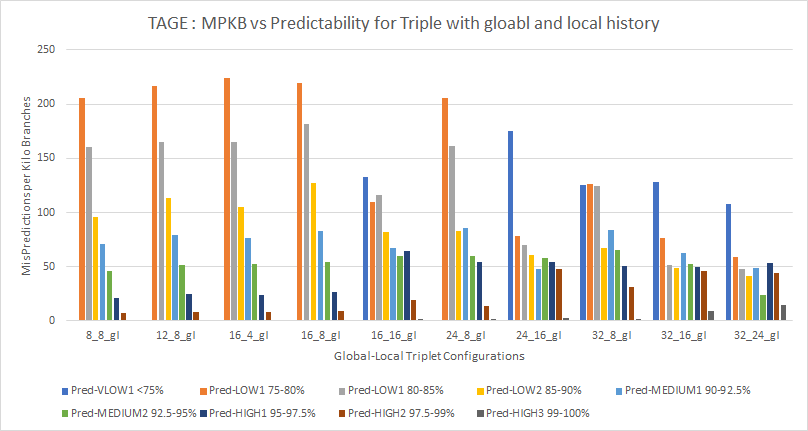}
\caption{TAGE: MPKB vs. Predictability in Global-Local Triplet's BWSET }
\label{GLpredcorrAS}
\end{figure}

The predictability in BWSET also correlates with MPKB of TAGE up to 16
bits of global and 8 bits of local history, shown in Figure
\ref{GLpredcorrAS}. This analysis indicates that a {16,8}
configuration is a sweet spot for global-local tuple based
characterization. Any other combinations of shorter history lengths
lead to not all unique triplets to be identified, thus decreasing
accuracy. Similarly, a longer history length leads to aliasing in
hardware constrained predictors, leading to reduced accuracy.

\begin{table}[h!]
\centering
\caption{Predictability Projection vs TAGE Accuracy - 24g Global Tuple configuration}
\begin{tabular}{|p{1.50cm}|p{1.75cm}|p{0.75cm}|p{1.50cm}|p{1.35cm}|}
\hline
\textbf{Category} & \textbf{Predictability Range} & \textbf{No of Traces} & \textbf{Predictability Projection} & \textbf{TAGE Accuracy Percentage} \\ \hline
Pred LOW1 & $ 75\% - 80\%$ & 8 & $77.52\%$ & $77.48\%$  \\ \hline
Pred LOW2 & $ 80\% - 85\%$ & 3 & $82.50\%$ & $82.07\%$  \\ \hline
Pred LOW3 & $ 85\% - 90\%$ & 151 & $88.42\%$ & $87.95\%$  \\ \hline
Pred MEDIUM1 & $ 90\% - 92.5\%$ & 200 & $91.24\%$ & $92.03\%$  \\ \hline
Pred MEDIUM2 & $ 92.5\% - 95\%$ & 292 & $93.75\%$ & $94.88\%$  \\ \hline
Pred HIGH1 & $ 95\% - 97.5\%$ & 337 & $96.34\%$ & $97.72\%$  \\ \hline
Pred HIGH2 & $ 97.5\% - 99\%$ & 402 & $98.25\%$ & $99.27\%$  \\ \hline
Pred HIGH3 & $ 99\% - 100\%$ & 1058 & $99.43\%$ & $99.84\%$  \\ \hline
\end{tabular}
\label{predcomparisonGlobal}
\end{table}

\begin{table*}[h!]
\centering
\caption{Predictability Projection vs TAGE and Perceptron accuracy - 16g\_8l Global Local Tuple configuration }
\begin{tabular}{|l|l|l|l|l|l|l|}
\hline
\multicolumn{3}{c|}{} & \multicolumn{2}{c|}{\textbf{Predictability Projection}} & \multicolumn{2}{c|}{\textbf{Accuracy Percentage}}\\ \hline
\textbf{Category} & \textbf{Predictability Bins} & \textbf{No of Traces} & \textbf{Mean} & \textbf{Median} & \textbf{Perceptron} & \textbf{TAGE} \\ \hline
Pred LOW1 & $ 75\% - 80\%$ & 7 & $77.96\%$ & $77.70\%$ & $78.11\%$ & $78.05\%$ \\ \hline
Pred LOW2 & $ 80\% - 85\%$ & 4 & $82.24\%$ & $82.04\%$ & $81.70\%$ & $81.82\%$ \\ \hline
Pred LOW3 & $ 85\% - 90\%$ & 101 & $89.09\%$ & $89.06\%$ & $87.12\%$ & $87.26\%$ \\ \hline
Pred MEDIUM1 & $ 90\% - 92.5\%$ & 187 & $91.14\%$ & $91.04\%$ & $91.66\%$ & $91.75\%$ \\ \hline
Pred MEDIUM2 & $ 92.5\% - 95\%$ & 273 & $93.97\%$ & $94.35\%$ & $94.56\%$ & $94.58\%$ \\ \hline
Pred HIGH1 & $ 95\% - 97.5\%$ & 320 & $96.44\%$ & $96.47\%$ & $97.33\%$ & $97.36\%$ \\ \hline
Pred HIGH2 & $ 97.5\% - 99\%$ & 384 & $98.47\%$ & $98.53\%$ & $99.10\%$ & $99.11\%$ \\ \hline
Pred HIGH3 & $ 99\% - 100\%$ & 1176 & $99.65\%$ & $99.68\%$ & $99.89\%$ & $99.90\%$ \\ \hline
\end{tabular}
\label{predcomparison}
\end{table*}

Modern branch predictors extract and learn branch context from sophisticated
branch histories where they can be selective about which histories to choose.
Our attempts focus on providing a simplistic first-order attempt at
characterizing branches based on traces' inherent branch behavior.  With a
workload characterization based on global-local tupples instead of just global
history tuple , the predictability projections are now more closer to the
actual accuracy numbers of both TAGE and Perceptron predictors in PRED\_MEDIUM
and PRED\_HIGH categories, as shown in Table \ref{predcomparison}. On the
other hand, both TAGE and Perceptron designs now lag a little more in
PRED\_LOW when compared to our projected predictability numbers, making this
category a possible area for improvement in the future.

\section{Conclusion}
\label{sec:conc}

In this paper, we propose two new branch behavior indicators for workloads -
branch working set and predictability, both of which are independently
determined irrespective of application domain the trace belongs to or
architecture of the branch predictor. We define a 3-way global-local tuple
based branch working set for each workload, {\em i.e.}, the total number of uniquely
identifiable high-frequency branch contexts present in that workload, which
dominates a predictor's accuracy. Based on the bias of the tuples in the
branch working set, we also define the workload's representative
predictability measure. We present an analysis framework that can extract
BWSET size and predictability parameters along with other branch-related
information from a set of 2,451 industry released traces having vibrant branch
behavior. Our work identifies that a global-local tuple consisting of branch
PC, 16 bits of global history, and 8 bits of local history provides a good
representation of branch context and is the most basic unit which a predictor
needs to learn.

We show that the branch working set's size and predictability correlate very
well with the achievable branch prediction accuracy for any modern branch
prediction scheme. Based on these newly identified branch characteristics, we
present a workload characterization to categorize traces into 7 branch
working-set size-based bins and 9 predictability based bins. We note that
traces belonging to smaller branch working sets and high predictability groups
have a higher prediction accuracy than the traces belonging to the higher
branch working set and low predictability categories, respectively. Our
analysis framework is capable of running two states of the art branch
prediction schemes, TAGE and Multiperspective perceptron. It leads us to
identify workload categories where both of these prediction schemes have a
scope of improvement.

We expect this work will help in identifying branch behavior in workloads and
estimate a modern predictor's accuracy conservatively. We also provide a big
pool of workload traces that are already characterized by their branch
behavior and can be used for a more robust comparison of current and future
branch prediction strategies.

\newpage

\bibliographystyle{unsrt} % <===========================================
\bibliography{myReference} % to use file created by filecontents ...

@misc{smith1983branch,
  title={Branch predictor using random access memory},
  author={Smith, James E},
  year={1983},
  month=jan # "~25",
  publisher={Google Patents},
  note={US Patent 4,370,711}
}

@inproceedings{eyerman2006characterizing,
  title={Characterizing the branch misprediction penalty},
  author={Eyerman, Stijn and Smith, James E and Eeckhout, Lieven},
  booktitle={2006 IEEE International Symposium on Performance Analysis of Systems and Software},
  pages={48--58},
  year={2006},
  organization={IEEE}
}

@inproceedings{yeh1991two,
  title={Two-level adaptive training branch prediction},
  author={Yeh, Tse-Yu and Patt, Yale N},
  booktitle={Proceedings of the 24th annual international symposium on Microarchitecture},
  pages={51--61},
  year={1991}
}

@article{yeh1992alternative,
  title={Alternative implementations of two-level adaptive branch prediction},
  author={Yeh, Tse-Yu and Patt, Yale N},
  journal={ACM SIGARCH Computer Architecture News},
  volume={20},
  number={2},
  pages={124--134},
  year={1992},
  publisher={ACM New York, NY, USA}
}

@inproceedings{yeh1993comparison,
  title={A comparison of dynamic branch predictors that use two levels of branch history},
  author={Yeh, Tse-Yu and Patt, Yale N},
  booktitle={Proceedings of the 20th annual international symposium on computer architecture},
  pages={257--266},
  year={1993}
}

@techreport{mcfarling1993combining,
  title={Combining branch predictors},
  author={McFarling, Scott},
  year={1993},
  institution={Technical Report TN-36, Digital Western Research Laboratory}
}

@inproceedings{lee1997bi,
  title={The bi-mode branch predictor},
  author={Lee, Chih-Chieh and Chen, I-CK and Mudge, Trevor N},
  booktitle={Proceedings of 30th Annual International Symposium on Microarchitecture},
  pages={4--13},
  year={1997},
  organization={IEEE}
}

@inproceedings{seznec2005analysis,
  title={Analysis of the o-geometric history length branch predictor},
  author={Seznec, Andre},
  booktitle={32nd International Symposium on Computer Architecture (ISCA'05)},
  pages={394--405},
  year={2005},
  organization={IEEE}
}

@article{seznec2006case,
  title={A case for (partially)-tagged geometric history length predictors},
  author={Seznec, Andr{\'e}},
  journal={Journal of InstructionLevel Parallelism},
  year={2006}
}

@inproceedings{jimenez2003fast,
  title={Fast path-based neural branch prediction},
  author={Jim{\'e}nez, Daniel A},
  booktitle={Proceedings. 36th Annual IEEE/ACM International Symposium on Microarchitecture, 2003. MICRO-36.},
  pages={243--252},
  year={2003},
  organization={IEEE}
}

@inproceedings{jimenez2005piecewise,
  title={Piecewise linear branch prediction},
  author={Jim{\'e}nez, Daniel A},
  booktitle={32nd International Symposium on Computer Architecture (ISCA'05)},
  pages={382--393},
  year={2005},
  organization={IEEE}
}

@inproceedings{jimenez2006controlling,
  title={Controlling the power and area of neural branch predictors for practical implementation in high-performance processors},
  author={Jim{\'e}nez, Daniel A and Loh, Gabriel H},
  booktitle={2006 18th International Symposium on Computer Architecture and High Performance Computing (SBAC-PAD'06)},
  pages={55--62},
  year={2006},
  organization={IEEE}
}

@inproceedings{jimenez2001dynamic,
  title={Dynamic branch prediction with perceptrons},
  author={Jim{\'e}nez, Daniel A and Lin, Calvin},
  booktitle={Proceedings HPCA Seventh International Symposium on High-Performance Computer Architecture},
  pages={197--206},
  year={2001},
  organization={IEEE}
}

@inproceedings{loh2005reducing,
  title={Reducing the power and complexity of path-based neural branch prediction},
  author={Loh, Gabriel H and Jim{\'e}nez, Daniel A},
  booktitle={Proceedings of the 5th Workshop on Complexity Effective Design (WCED5)},
  pages={1--8},
  year={2005},
  organization={Citeseer}
}

@article{jimenez2016multiperspective1,
  title={Multiperspective perceptron predictor},
  author={Jim{\'e}nez, D},
  journal={Championship Branch Prediction (CBP-5)},
  year={2016}
}

@article{jimenez2016multiperspective2,
  title={Multiperspective perceptron predictor with TAGE},
  author={Jim{\'e}nez, D},
  journal={Championship Branch Prediction (CBP-5)},
  year={2016}
}

@article{seznec201164,
  title={A 64 kbytes ISL-TAGE branch predictor},
  author={Seznec, Andr{\'e}},
  journal={JWAC: Championship Branch Prediction},
  year={2011}
}

@article{seznec2014tage,
  title={Tage-sc-l branch predictors},
  author={Seznec, Andr{\'e}},
  journal={JWAC4: Championship Branch Prediction (CBP-4)},  
  year={2014}
}

@article{seznec2016tage,
  title={Tage-sc-l branch predictors again},
  author={Seznec, Andr{\'e}},
  journal={Championship Branch Prediction (CBP-5)},
  year={2016}
}

@article{tarjan2005merging,
  title={Merging path and gshare indexing in perceptron branch prediction},
  author={Tarjan, David and Skadron, Kevin},
  journal={ACM transactions on architecture and code optimization (TACO)},
  volume={2},
  number={3},
  pages={280--300},
  year={2005},
  publisher={ACM New York, NY, USA}
}

@article{chang1996branch,
  title={Branch classification: a new mechanism for improving branch predictor performance},
  author={Chang, Po-Yung and Hao, Eric and Yeh, Tse-Yu and Patt, Yale},
  journal={International Journal of Parallel Programming},
  volume={24},
  number={2},
  pages={133--158},
  year={1996},
  publisher={Springer}
}

@inproceedings{haungs2000branch,
  title={Branch transition rate: a new metric for improved branch classification analysis},
  author={Haungs, Michael and Sallee, Phil and Farrens, Matthew},
  booktitle={Proceedings Sixth International Symposium on High-Performance Computer Architecture. HPCA-6 (Cat. No. PR00550)},
  pages={241--250},
  year={2000},
  organization={IEEE}
}

@inproceedings{yokota2008potentials,
  title={Potentials of branch predictors: From entropy viewpoints},
  author={Yokota, Takashi and Ootsu, Kanemitsu and Baba, Takanobu},
  booktitle={International Conference on Architecture of Computing Systems},
  pages={273--285},
  year={2008},
  organization={Springer}
}

@article{de2016linear,
  title={Linear branch entropy: Characterizing and optimizing branch behavior in a micro-architecture independent way},
  author={De Pestel, Sander and Eyerman, Stijn and Eeckhout, Lieven},
  journal={IEEE Transactions on Computers},
  volume={66},
  number={3},
  pages={458--472},
  year={2016},
  publisher={IEEE}
}

@article{chen1996analysis,
  title={Analysis of branch prediction via data compression},
  author={Chen, I-Cheng K and Coffey, John T and Mudge, Trevor N},
  journal={ACM SIGPLAN Notices},
  volume={31},
  number={9},
  pages={128--137},
  year={1996},
  publisher={ACM New York, NY, USA}
}

@misc{cbp5,
  title={Championship Branch Prediction (CBP-5)},
  year={2016},
  howpublished = {\url{https://www.jilp.org/cbp2016/}}
}

@misc{cvp,
  title={Championship Value Prediction, 2018},
  year={2018},
  howpublished = {\url{https://www.microarch.org/cvp1/}}
}

\end{document}